\documentstyle[itzidef]{mn}

\input epsf

\title[Multicolour imaging of $z$= 2 QSO hosts]{Multicolour imaging of 
$z$ = 2 QSO hosts}

\author[I. Aretxaga, R.J. Terlevich \& B.J. Boyle]
       {Itziar Aretxaga$^1$, Roberto J. Terlevich$^2$,
B.J. Boyle$^3$\\
$^1$ Max-Planck-Institut f\"ur Astrophysik, Karl Schwarzschildstr. 1,
Postfach 1523, 85740 Garching, Germany\\
$^2$ Royal Greenwich Observatory, Madingley Road, Cambridge CB3 0EZ, U.K.\\
$^3$ Anglo-Australian Observatory, PO Box 296, Epping, NSW 2121 Australia}
\date{\large M\"unchen 97 July}

\begin{document}

\maketitle

\begin{abstract}
We present multicolour images of the hosts of three $z = 2$ QSOs
previously detected in $R$-band by our group. The luminosities,
colours and sizes of the hosts overlap with those of actively
star-forming galaxies in the nearby Universe.  Surface brightness
radial profiles over the outer resolved areas roughly follow either a
$r^{1/4}$ or an exponential law. These properties give support to the
young host galaxy interpretation of the extended light around QSOs at
high-redshift.  The rest-frame UV and UV-optical colours are
inconsistent with the hypothesis of a scattered halo of light from the
active nucleus by a simple optically-thin scattering process produced
by dust or hot electrons.  If the UV light is indeed stellar, star
formation rates of hundreds of solar masses per year are implied, an
order of magnitude larger than in field galaxies at similar redshifts and
above.  This might indicate that the QSO phenomenon (at least the
high-luminosity one) is preferentially acompanied by enhanced galactic
activity at high-redshifts.

\end{abstract}



\section{Introduction}

Recent deep-imaging studies have discovered a large population of
young galaxies at redshifts $z>0.3$, about 3\%\ being at $z \gsim 2.5$
for $R\lsim25$ (e.g. Steidel et al. 1996a). 
The galaxies have properties very similar
to those of present day star-forming dwarf galaxies or H~II galaxies,
but with star formation rates SFR$\,\lsim 30\,$\Msun /yr they are not
particularly luminous objects (Guzm\'an et al. 1997; Phillips et al. 1997).
The field population of high-$z$ galaxies has been associated with
early counterparts of disks (Wolfe et al. 1995), small spheroids of
present-day $L \gsim L^*$ galaxies (Steidel et al. 1996a) and/or
with a collection of star-forming knots being merged into 
galaxies (Lowenthal et al. 1997).  On the
other hand, detailed studies of the intrinsic properties of nearby
cluster ellipticals indicate that their stellar population was formed
at $z>2$ (Bower, Lucey \& Ellis 1992).  One long standing view is that
these were systems formed monolithically in an early burst of star
formation (e.g. Larson 1974), and have been passively evolving since
then. Thus, for a first rank elliptical we expect SFR$\,\sim
{10^{12}\Msun}/{10^9 {\rm yr}}$.  Several attempts to find the early
counterparts of these galaxies in large areas of the sky have been
conducted (see Pritchet 1994 for a review) but to date just a handful
of candidates that still need spectroscopic confirmation have been
found (e.g. Pahre et al. 1997).  Where are these bright objects hiding?  In
galaxy formation models based on the hierarchical merging of dark
matter halos the majority of large spheroids are formed at lower
redshifts (Kauffmann, White \& Guiderdoni 1993), so that the
present-day large systems were previously broken-up into small fainter
pieces of a yet not completely assembled whole.  Although they are
an attractive solution, it still needs to be demonstrated whether these
models can reproduce the intrinsic properties of massive ellipticals,
such as their large nuclear metal content, the thinness of the
fundamental plane, the small scatter of the colour-luminosity
relation, the large colour/metal gradients, etc. (but see Kauffmann \&
Charlot 1997 for first positive results on the colour-luminosity
relation).  There is also a good possibility that some large spheroids
may have gone undetected as the host galaxies of bright QSOs.  Two
pieces of observational evidence support this hypothesis.  First, a
significant number of nearby bright QSOs are associated with luminous
ellipticals (McLeod \& Rieke 1995; Disney et al. 1996; Taylor et
al. 1996).  Secondly, many of the properties of high-$z$ QSOs are
consistent with their association with the cores of early luminous
spheroids. Among them, a) the high metal content of the Broad Line
Region of high-$z$ QSOs (Hamman \& Ferland 1993); b) the detection of
large masses of dust in $z\approx 4$ QSOs (Isaak et al. 1994); and c)
the correspondence of the observed luminosity function of QSOs and
ellipticals (Terlevich \& Boyle 1993 , Haehnelt \& Rees 1993) do
suggest that QSO activity might be occurring in the central regions of
young massive ellipticals.

All these points are reinforced by the recent discovery that about 5
to 20\% of the luminosity of both radio-loud and radio-quiet QSOs at
$z \approx 2$ arises from extended structures of FWHM$\,\approx
1-2$~arcsec (Lehnert et al. 1992, Aretxaga, Boyle \& Terlevich 1995,
Hutchings 1995).  Unraveling the nature of the extensions around QSOs
at these redshifts is of major importance since they could be
revealing the hidden signature of the elusive early large spheroid
population.

In this paper we focus our attention on the multicolour properties of
the hosts of three $z \approx 2$ QSOs previously detected in $R$-band
by Aretxaga, Boyle \& Terlevich (1995; hereinafter ABT95), in an
attempt to characterize the emission mechanism. In section~2 we
describe the dataset.  In section~3 we describe the method of finding
the hosts, and their analysis. In section~4 we discuss the nature of
the hosts on the basis of their colours, luminosities, sizes and
profiles. In section~5 we summarize our main conclusions.


\ifoldfss
  \section{Data set: Observations and Reduction}
\else
  \section[]{Data set}
\fi
  
  The three QSOs studied here were previously detected to be extended
in the $R$-band imaging program of ABT95.  They comprise the three
extended objects in their four target list.  The QSOs were originally
selected from the Ver\'on-Cetty and Ver\'on (1994) catalogue with the
only condition of belonging to a narrow redshift-luminosity band ($1.8
\lsim z \lsim 2.2$, $M_B \lsim -28$ mag for \Ho50\ and \qo0p5), and
lying close in projection ($20 \lsim \theta \lsim 50$~arcsec) to stars
of similar brightness in order to define a reliable 
point-spread-function (PSF).
The QSOs, two radio-quiet and one radio-loud, can thus be said to be
representative of the high-luminosity end of the QSO population at the
epoch when QSOs were most abundant (e.g. Boyle et al. 1991).

 Our observations (including those already published in ABT95)
are summarized in Table~1, and detailed in sections 2.1 and 2.2.
 \begin{table*}
 \begin{minipage}{140mm}
 \begin{center}
  \caption{Summary of observations}
  \begin{tabular}{lcccrlllc}
   Name     &  $M_B^*$ & $z^*$ & Filter & Exposure & seeing$^{\dag}$ &
Telescope & Detector & Date \\  
            &   &   &   & (seconds) &(arcsec)  & & &   \\
1630.5$+$3749   & $-28.3$ & 2.037 & $R$ & 10500 & 0.6 & WHT 4.2m & TEK1
& 1994 Aug  \\
		&	  &	  & $I$ & 4800 & 0.75 & WHT 4.2m & TEK5
& 1995 Sep \\
	   	&	  &	  & $K'$	& 13320 & 1.0 & Calar Alto
3.5m & MAGIC & 1995 Aug\\
PKS 2134$+$008$^{\ddag}$  
& $-29.6$ & 1.936 & $R$ & 5850  & 0.6 & WHT 4.2m & TEK1
& 1994 Aug  \\
		&	  &	  & $I$ & 4800  & 0.9 & WHT 4.2m & TEK5
& 1995 Sep \\
	   	&	  &	  & $K'$	& 5220 & 1.1 & Calar Alto
3.5m & MAGIC & 1995 Aug\\

Q 2244$-$0105   & $-28.6$ & 2.040 & $R$ & 5400 & 0.7 & WHT 4.2m & TEK1
& 1994 Aug  \\
		&	  &	  & $R$ & 3600  & 0.9 & WHT 4.2m & TEK5
& 1995 Sep \\
		&	  &	  & $I$ & 3600  & 0.7 & WHT 4.2m & TEK5
& 1995 Sep \\
	   	&	  &	  & $K'$	& 7670 & 1.0 & Calar Alto
3.5m & MAGIC & 1995 Aug\\

 \end{tabular}
\end{center}
$^*$ $M_B$ and $z$ from V\'eron--Cetty and V\'eron (1994).\\
$^{\dag}$  Stellar FWHM measured on co-added frame.\\
$^{\ddag}$ radio-loud QSO\\
\end{minipage}
\end{table*}

\subsection{Optical Imaging}

The principal optical observations in this study were performed at the
f/11 Auxiliary Port of the 4.2m William Herschel Telescope
(WHT\footnote{The William Herschel and the Issac Newton
Telescopes are operated on the island of La Palma by the Royal
Greenwich Observatory in the Spanish Observatorio del Roque de los
Muchachos of the Instituto de Astrof\'{\i}sica de Canarias}) in La
Palma.  The 1024x1024 pixel CCD TEK chips in this configuration give a
spatial resolution of $0.105$~arcsec/pixel over an unvignetted
$1.8\,$arcmin diameter field. The CCD was read out in QUICK mode.

Harris $R$-band images for all three QSOs were obtained in August 1994 
under non-photometric conditions. The reduction and analysis
of this data set was already discussed in ABT95, and does not  differ
from that performed in the present analysis. 
In 1995 September 3/4 we acquired Harris $I$-band images of the three
QSO fields, and a second set of $R$-band observations for
Q~2244$-$0105.  The observations were also obtained under
non-photometric conditions.

For each observation, the total integration time of each QSO was divided in
600s exposures to avoid saturation of the QSOs and stars. 
Individual frames were offset from
one another by $\sim 10\,$arcsec to allow discrimination of bad pixels.

The data were reduced with the IRAF\footnote{IRAF is distributed by
the National Optical Astronomy Observatories, which are operated by
the Association of Universities for Research in Astronomy, Inc., under
cooperative agreement with the National Science Fundation} software
package. Each QSO frame was first bias subtracted and then
flat-fielded using a sky flat-field.  The sky-flat field was computed
for each night from the median of the frames of all QSO fields, after
bright stars, galaxies and QSOs had been masked from the data.

Photometric zero-points for the $R$ and $I$ band data on the fields
around 1630.5$+$3749 and PKS~2134$+$008 were obtained based on
observations made with the 2.5m Issac Newton Telescope (INT$^\star$)
during service time on 1996 September 19.
Magnitudes were calculated through circular apertures which reached
the asymptotic value of the circular aperture growth, typically about
3 to 4 arcsec in radius.

\subsection{Near-infrared Imaging}

The near-infrared imaging was conducted at the 3.5m
telescope of the German-Spanish Astronomical Center
on Calar Alto 
\footnote{The German-Spanish Astronomical Center on Calar Alto is
operated by the Max-Plack-Institute for Astronomy, Heidelberg, jointly
with the Spanish National Comission for Astronomy} with MAGIC, the
256x256 pixel NICMOS3 CCD array.  The f/10 configuration in
high-resolution mode gives a spatial resolution of 0.32~arcsec/pixel
over a square 82x82~arcsec field. The chip was read out in the standard
reset.read.read mode.

On 1995 August 5/6/7 we obtained $K'$ images of the three QSO fields
under photometric conditions The total integration times for the QSO
fields were split into 2s units to avoid sky saturation, and stacked
in 1 min frames. As with the optical imaging, individual frames were
offset from one another by $\sim 10\,$arcsec, making sure that both
QSOs and nearby PSF stars were included in the field-of-view of the
detector.

The data were reduced with the IRAF software package.  Sky and
dark-current were subtracted in each 
frame using the median of six
unregistered frames of the same field obtained just before and after
each observation, and scaled up to the mean value of source-free
patches of the sky. The frames were then flat-fielded using a
dark-current corrected exposure of the uniformly illuminated
dome. Each QSO field was co-registered and co-added using the
centroids of the brightest stars and QSOs.  Flux calibration was
performed through a set of standard stars from the list of Elias et
al. (1982) throughout the nights. Integrated magnitudes for all the
objects were obtained through circular aperture photometry that
reached the asymptotic values of the circular aperture growth.

\subsection{Near-infrared Spectroscopy}

A 5 min integration spectrum of PKS 2134$+$008 and a 10 min
integration spectrum of Q 2244$-$0105 were also obtained for our
program during service time with the CGS4 spectrograph at the 3.8m UK
Infrared Telescope (UKIRT\footnote{The United Kingdom Infrared
Telescope is operated by the Joint Astronomy Center on behalf of the
UK Particle Physics and Astronomy Research Council}) on Mauna Kea. The
observations were acquired with spectral resolution R=550 and spectral
coverage 1.45--2.05$\mu$, in order to contain the rest-frame
[O~III]\ldo{5007} to \Ha\ region. The spectra were flux and wavelength
calibrated at the observatory following their standard procedures.


\ifoldfss
  \section{Data analysis}
\else
  \section[]{Data analysis}
\fi

\subsection{Optical Imaging}

\subsubsection{PSF subtraction}

The procedure used to fit and subtract the stellar PSF from the images
was described in ABT95 in detail. Briefly, it involved the definition
of a two-dimensional PSF from the brightest stellar companion to the
QSO (marked in Fig.~1 of ABT95), and the subtraction of the 2D-profile
from the QSO and nearby stars, following three recipes: (1) flux-scaled
subtraction; (2) subtraction that yields zero flux in the center of
the QSO; (3) subtraction that yields a smooth flat-top profile with no
depression in the center. Criteria (2) and (3) were evaluated for each
frame after smoothing the residuals of the subtraction with a Gaussian
filter of $\sigma = 1$~pixel.  This procedure provided stability to
the solution found from frame to frame.  The frames were then
registered and co-added using the centroids of stars and QSOs in the
original unsubtracted frames.  Only non-smoothed frames were used in
the co-addition.

\begin{figure*}
    \cidfig{5.0in}{25}{160}{579}{696}{ 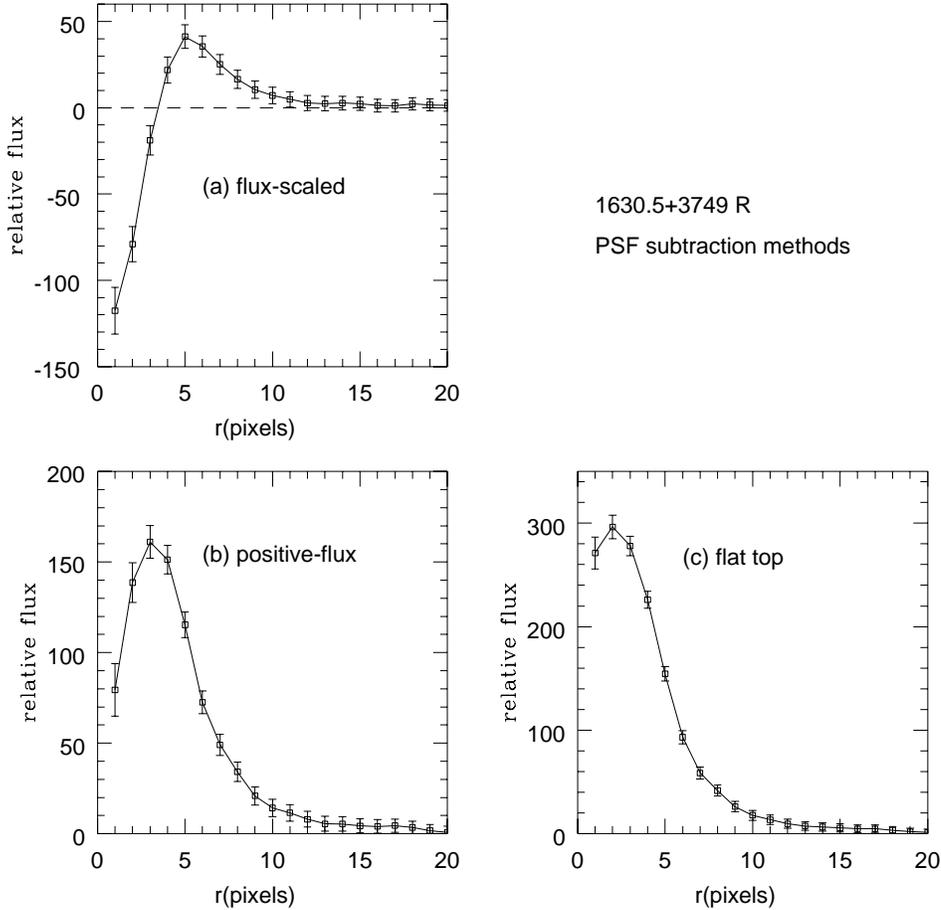}
    \caption{Radial profiles of the solutions found by performing PSF 
subtractions on the $R$-band frames of 
1630.5+3749. Error bars represent the scatter around the mean value,
and do not include the subtraction error.
{\bf (a)} flux-scaled subtraction; {\bf (b)} subtraction to obtain 
zero flux in the position of the QSO centroid; 
{\bf (c)} subtraction to obtain a smooth flat-top profile
}
\end{figure*}

Flux-scaled subtraction typically produces solutions with
negative flux in the center, defined as the centroid of the original
QSO position, surrounded by a ring of positive flux (see Fig.~1a).
For the three QSOs the flux-scaled PSF subtraction is a solution that
is inconsistent with QSOs and PSF stars having the same spatial
profiles.  The QSO residuals left after the QSO$-$PSF subtraction depart from
the subtraction error ($\sigma$) by over $3\sigma$ in both 
$R$ and $I$-band images, 
with the only exception being
the $I$-band image of Q~2244$-$0105 ($2\sigma$).  As a check, bright nearby
PSF-subtracted stars leave residuals around $1\sigma$
(see also Figure~2 in ABT95
for a graphic comparison of stellar and QSO residuals in the $R$-band frames).

Fig. 1b and 1c show typical radial profiles for the two other
subtraction criteria. These will be regarded as a lower limit and 
a good approximation to the actual profile of the hosts.

Fig.~2 shows contour plots of the central-zero-flux 
subtraction solutions. The putative hosts are roundish and show no obvious
signs of disrupted morphologies for our limiting PSF-subtraction 
backgrounds ($\mu_R \approx 25.0$ to 25.5~mag, $\mu_I \approx 23.0$ to
23.5~mag at $1\sigma$
per pixel, evaluated from the sky in the PSF-subtracted area
around the QSOs).

\begin{figure*}
    \cidfig{7.0in}{35}{69}{562}{755}{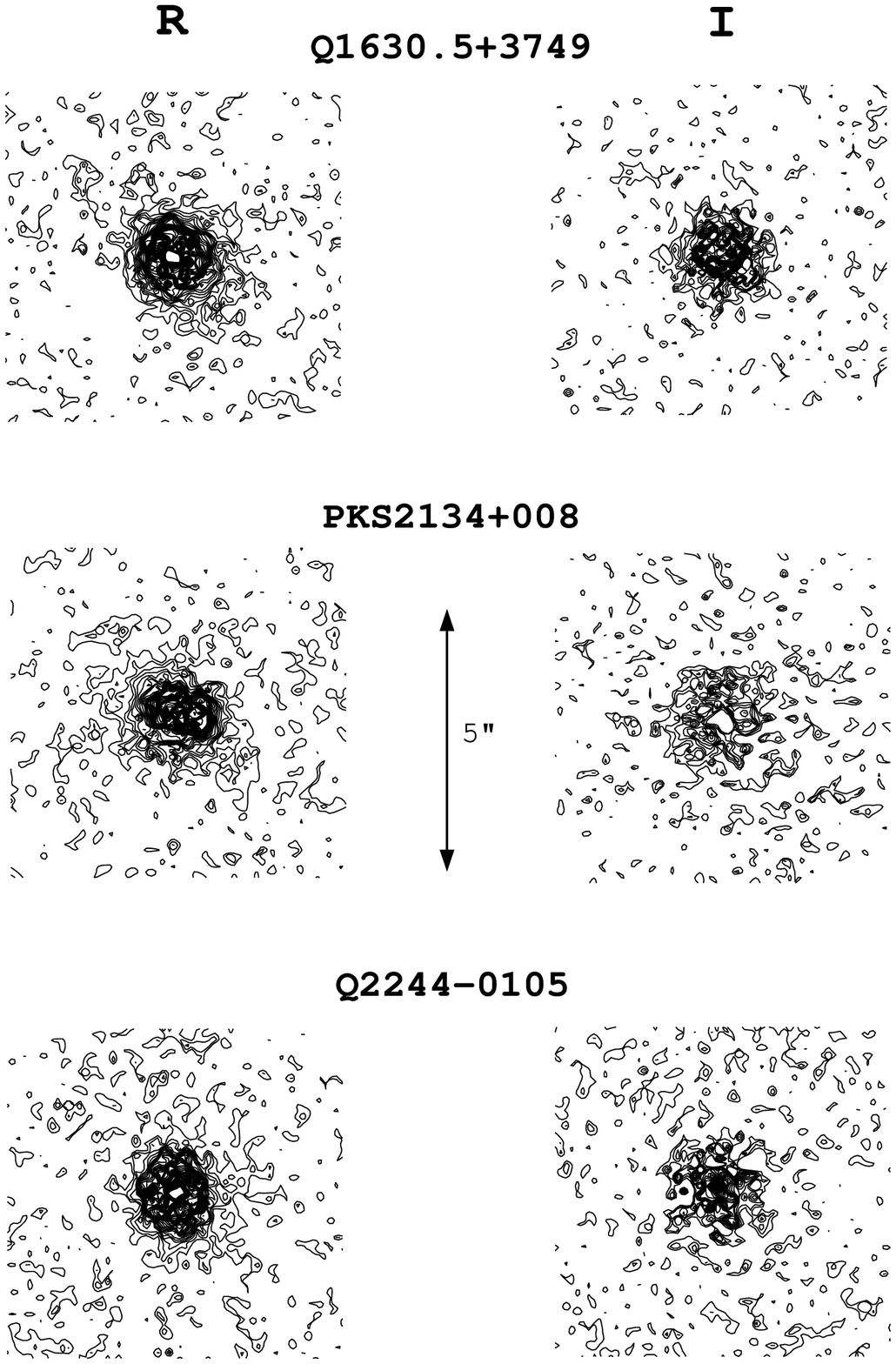}
    \caption{$R$ and $I$-band contour plots of the solutions found
after performing a PSF subtraction that yields zero flux in the 
centroid of the QSO. The minimum contour level was set at $1\sigma$ of the sky
noise after subtraction of the PSF. Further levels indicate $1\sigma$ 
increments over the minimum level. North is to the left, and East is down.
}
\end{figure*}

\subsubsection{Accuracy of PSF subtractions}

To assess the accuracy of the PSF subtractions and estimate the errors
on the derived properties of the hosts, we have 
constructed a set of simulated QSOs by superposing observed stellar PSFs on
compact faint field galaxies and also on compact simulated galaxies 
(FWHM$\approx 0.8 - 1.2$~arcsec). The simulated galaxies were introduced
in our images with the package ARTDAT of IRAF 
with the same Poissonian noise as that of the observations.
Nuclear components using PSF models derived from field stars and
containing  90, 95, 97, and
99 per cent of the total luminosity of the simulated QSOs were superposed
on the galaxies,
and the PSF subtraction was then attempted with the same 
PSF models. The results for two of these galaxies are shown in Fig.~3. 

As one would assume intuitively, the simulations show that the larger
the size and the larger the contribution of the host to the total
luminosity of the QSO, the easier it is to recover its true
luminosity, radius and profile. The simulations also show that
flat-top subtractions tend to underestimate the total luminosity of
the hosts.  This effect is actually introduced by design through the
smoothing performed in order to evaluate whether a flat regime had
been attained.  This shows that although flat-top PSF subtractions
yield better estimates for the host properties than central-zero-flux
subtractions, they are still conservative. The biggest departures from
the true profiles (about a factor of 2) are expected in the inner
stellar FWHM scale for the range of galaxy compactness explored. These
discrepancies will be even greater for even more compact galaxies.  As
an example, from the circular aperture curves of growth of the most
compact galaxies in our simulations (FWHM$\approx 0.8-1$~arcsec), we
derive that the flux of the hosts can be measured within better than a
factor of 2 if the host contribution is between 1 and 5 per cent of
the total QSO luminosity, and increasingly better for larger
luminosities and sizes. The effective radii of the hosts can be
derived with relative errors between 20 and 10 per cent, for
contributions of the galaxy between 3 and 5 per cent of the QSO
luminosity.

\begin{figure*}
    \cidfig{5.0in}{29}{154}{582}{449}{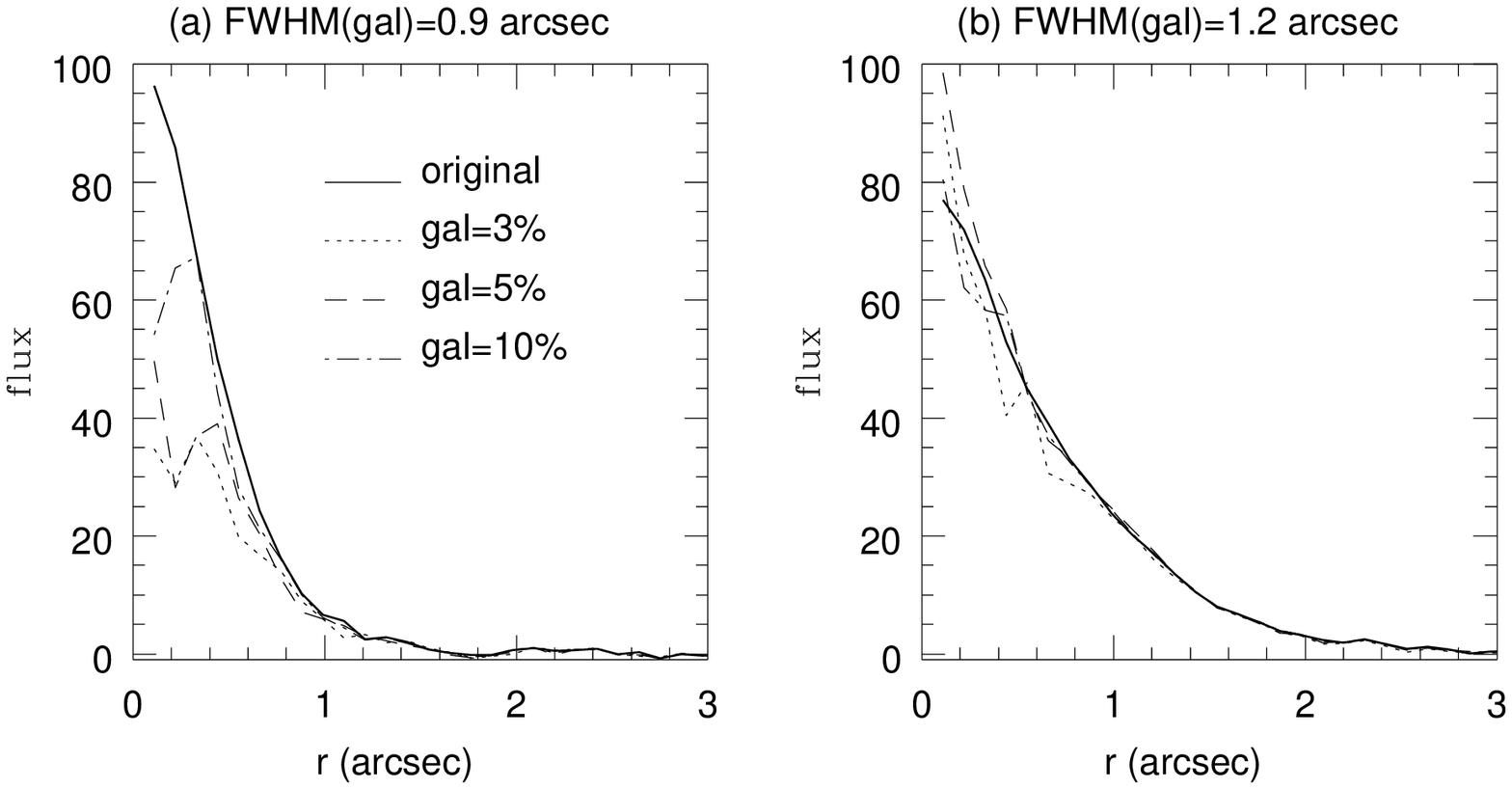}
    \caption{Results of flat-top PSF subtractions on simulated QSOs that
use galaxies of (a) FWHM$=0.9$ and (b) FWHM$=1.2$~arcsec, with
nuclear profiles of FWHM$=0.8$~arcsec.
The solid lines represent the
original profiles of the galaxies before superposing the unresolved components.
Dashed lines represent the recovered profiles of the galaxies
after PSF subtraction.
For the simulations presented here, the galaxies comprised
3, 5, and 10 per cent of the total luminosity of the simulated QSOs.
}
\end{figure*}

\subsubsection{Properties of the hosts}

Integrated magnitudes, effective radii, and radial profiles of the flat-top 
PSF subtracted hosts were computed. The results are summarized in
Table~2 and Fig.~4.

\begin{figure*}
    \cidfig{6.0in}{43}{159}{574}{700}{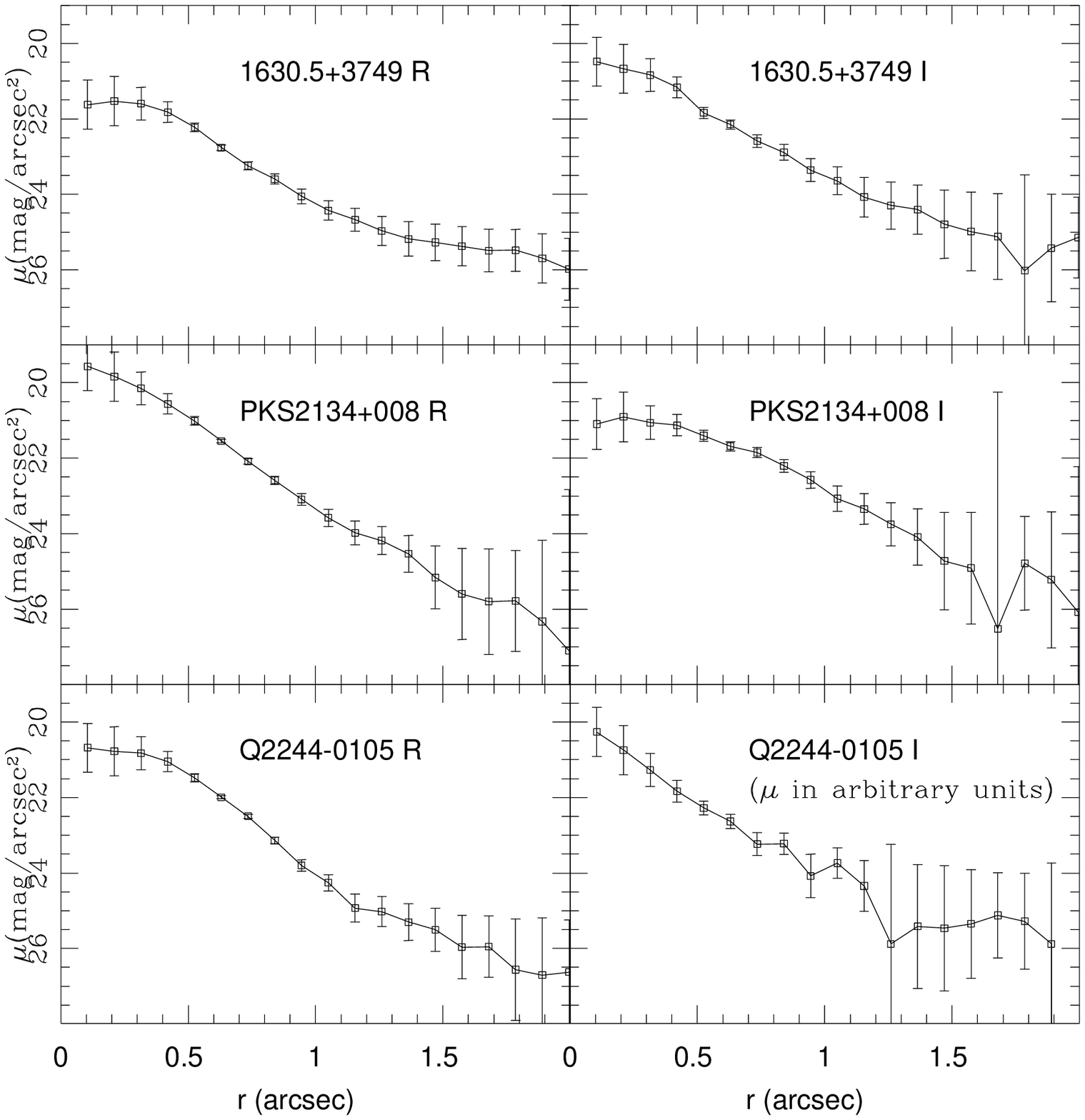}
    \caption{Radial profiles of the hosts derived from flat-top PSF 
subtractions. The error bars in the inner regions are dominated by 
the error in the subtractions, estimated from the simulations, and in the 
outer regions by sky subtraction.
The $I$-band profile of Q 2244$-$0105 is not flux-calibrated. 
}
\end{figure*}

 \begin{table*}
 \begin{minipage}{140mm}
 \begin{center}
  \caption{Integrated optical magnitudes and scale-lengths of hosts and QSOs.
The columns give: name of the QSO; filter; 
$m$(QSO) magnitude of QSO$+$host; $m_1$ magnitude of the hosts found with 
central zero-flux PSF subtraction; $m_2$ magnitude of the hosts found with 
flat-top PSF subtraction ; $D_{eff}$ effective diameter of the hosts 
found with flat-top PSF subtraction;
seeing of the co-added frames, defined as the FWHM of the stellar profiles. 
}
  \begin{tabular}{lccccccr}
   Name     & Filter & $m$(QSO) & $m_1$ & $m_2$ & $D_{eff}$ & seeing \\
            &        & (mag)    & (mag) & (mag) & (arcsec)  & (arcsec) \\
1630.5$+$3749  &  $R$  &  $18.34 \pm 0.12$  &  22.0 & 21.3 &  0.9 &   0.6\\
	       &  $I$  &  $17.55 \pm 0.10$  &  20.6 & 20.2 &  1.0 &   0.75
\\
PKS 2134$+$008 &  $R$  &  $16.78 \pm 0.10$  &  20.8 & 20.2 &  0.95 & 0.6 \\
	       &  $I$  &  $16.19 \pm 0.10$  &  19.7 & 19.4 &  1.2 & 0.9 \\
Q 2244$-$0105  &  $R$  &  $17.5 \pm 0.30 \  ^{\dag}$   &  $m$(QSO) $+3.4$ &  
$m$(QSO) $+2.3$ &   0.8 & 0.7 \\
	       &  $I$  &  	---	    &  $m$(QSO) $+3.6$ &
$m$(QSO) $+3.3$ &   1.0 & 0.7 \\
 \end{tabular}
\end{center}
$^{\dag}$  Flux calibration based on APM Northern Sky Catalogue.
\end{minipage}
\end{table*}

%
%

 \begin{table*}
 \begin{minipage}{140mm}
 \begin{center}
  \caption{Colours of nuclei and hosts.
The columns give: name of the QSO; colour; colour of nucleus; 
total colour of host (1) from the flat-top
 subtractions; colour of the host (2) in the outer 1 to 3 arcsec. 
}
  \begin{tabular}{llcccc}
   Name     & Colour & colour(AGN) & colour(host)$^1$ &
 colour(host)$^2$ \\
            &        & (mag)    & (mag) & (mag) \\
1630.5$+$3749  &  $R-I$  &  $0.8 \pm 0.1$  & $1.1 \pm 0.1$ &
$1.0 \pm 0.15$ \\ 
	       &  $R-K$  &  $2.5\pm0.1$  &  --- & $3.3\pm 0.1$ \\
PKS 2134$+$008 &  $R-I$  &  $0.6\pm0.1$  &  $0.8\pm0.1$ &
 $1.1\pm0.1$ \\
	       &  $R-K$  &  $2.1 \pm 0.1$  &  --- & $\lsim 3.3$\\
Q 2244$-$0105  &  $R-I$  &  --- & --- & --- \\
	       &  $R-K$  &  $3.4\pm0.3$ & --- &  $\lsim 3.3$\\
 \end{tabular}
\end{center}
\end{minipage}
\end{table*}

The magnitudes of the QSO systems (i.e. nuclear $+$ host light)
are calibrated using our own observations, except for the $R$-band image of
Q 2244$-$0105, which was calibrated using the crude zero-points of ABT95
based on the APM Northern Sky Catalogue
(Irwin, Maddox \& McMahon 1994). 
The difference between these zero-points and
our calibration for the objects in the other two QSO fields 
is within the error bars of both measurements, but 
our new calibration is typically 0.4~mag fainter. 

Flat-top PSF subtractions yield hosts that comprise
between 5 and 12 per cent of the total luminosity of the QSO$+$host systems.

Table~3 contains the colours of the nuclei and host galaxies.  The
colours of 1630.5$+$3749 ($R-I = 0.8 \pm 0.1$~mag) and
PKS 2134$+$008 ($R-I = 0.6 \pm 0.1$) are within the range of observed
colours of
QSOs at that redshift (V\'eron \& Hawkins 1995).  These
colours are somewhat bluer than those derived for the hosts: $R-I \sim
1.1\pm 0.1$~mag and $R-I \sim 0.8 \pm 0.1$~mag, respectively.  
However, the errors
associated with the PSF subtraction are usually larger than the
calibration errors, and could easily account for this discrepancy
between QSO and host colours.  In order to obtain a more reliable
value of the colours of the hosts, we have integrated the total luminosity
enclosed between two concentric circular apertures of radius
0.7~arcsec and 3~arcsec.  The outer aperture reaches the plateau of
the aperture growth curves, and the inner one neglects the part of the
hosts more affected by subtraction errors (see Fig.3). For the host of
1630.5$+$3749 we find $R-I = 1.03$ and for that of PKS 2134$+$008
$R-I=1.1$, redder by 0.24 and 0.51~mag, respectively, than the total light of the
QSOs.  The colours remain similar within rms fluctuations of 0.15 and
0.06~mag respectively when the
inner aperture is set from 0.7~arcsec to 1~arcsec, indicating that the
outermost regions of the hosts of PKS 2134$+$008  and
1630.5$+$3749 are redder than their nuclei.

\subsection{Infrared Imaging}

A two-dimensional PSF subtraction technique similar to that described
above for the optical images was also performed in the infrared
frames. The only difference is that the individual frames were
registered and co-added before performing the subtraction. The
residuals left in the QSO positions were in all cases negligible.

Radial profile comparison of QSOs and PSF stars in the field show that
the differences between them are indeed small, and located in the
outermost regions, where the sky subtraction is critical. Fig.~5 shows
the radial profiles normalized to the first computed ring at
0.32~arcsec for QSOs and stars. Only 1630.5$+$3749 has a marginal
excess above the stellar profile at 2 to 4~arcsec from the center.
The excess does not disappear even after allowance for shifts in the
sky level.

The colours of the QSOs 1630.5$+$3749 ($R-K = 2.5 \pm 0.1$~mag), PKS
2134$+$008 ($R-K = 2.1 \pm 0.1$~mag) and Q 2244$-$0105 ($R-K\approx
3.4 \pm 0.3$~mag) are similar to those measured in other 
QSOs at these redshifts (Hewett, priv. communication).  
If we assume that the excess
present in 1630.5$+$3749 is real, the surface brightness colour over
the 2--3~arcsec region is $R - K = 3.3 \pm 0.2$~mag, redder by about
0.8~mag than the observed colour of the QSO.

The surface brightness limits for the $K$-band images are between
22.1 and 21.8~mag/arcsec$^2$ at $1\sigma$ per pixel. The limits for
$3\sigma$ detections of point-like objects with a typical seeing-disk range
between 20.8 and 21.1~mag.  These limits are similar to those obtained
in a previous $K$-band imaging program of radio-quiet QSOs at $z
\approx 2$ that yielded non-detections (Lowenthal et al. 1995).

From the $K$-band limits we can set the colours of the hosts
of PKS 2134$+$008 and Q 2244$-$0105 to be
$R-K \lsim 3.3$~mag, consistent with those of the 
unresolved nuclear light.

\begin{figure*}
    \cidfig{5.0in}{43}{159}{574}{700}{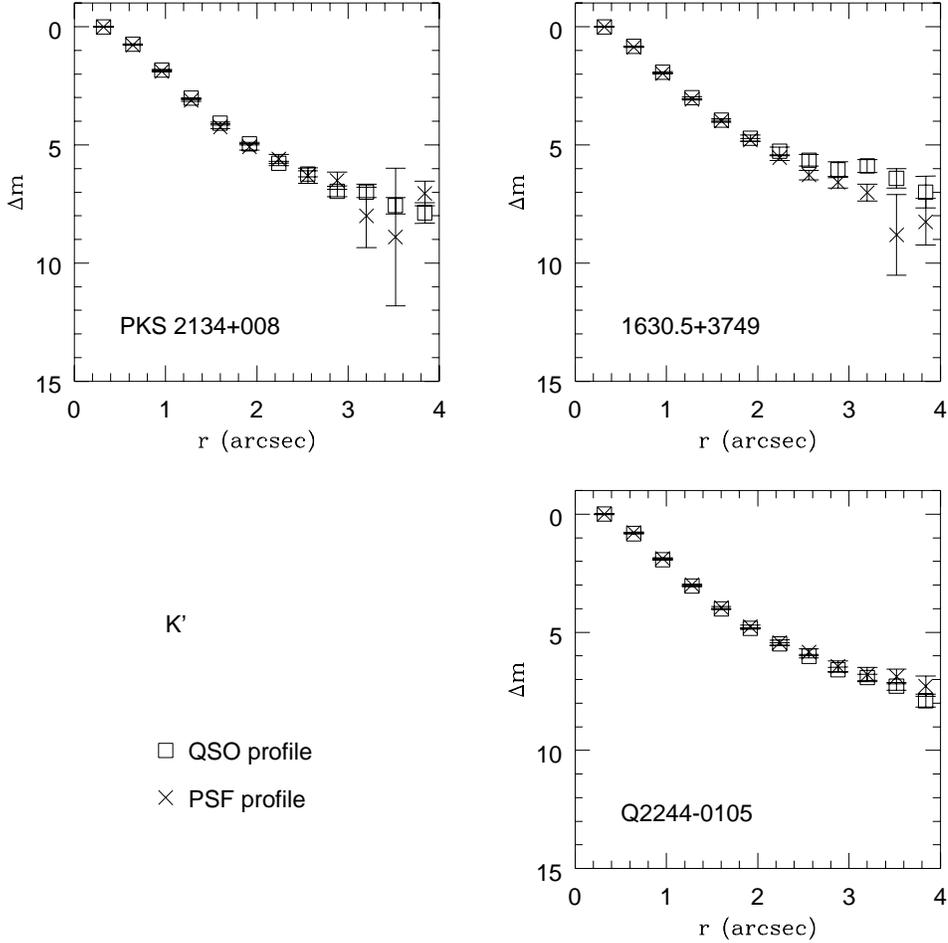}
    \caption{$K'$-band radial profiles of QSOs and stars in the field, 
normalized to their central flux.
}
\end{figure*}

\ifoldfss
  \section{Discussion: the nature of the hosts}
\else
  \section[]{Discussion: the nature of the hosts}
\fi

The hosts we have detected have luminosities that contribute between 5
and 12\% of the luminosity of the QSOs (nucleus$+$extension) both in
$R$ and $I$-bands (\ldo{2300} and \ldo{2800} rest-frame).  If the
marginal excess over the stellar profile of 1630.5$+$3749 is real, the
$K$-band (\ldo{7600} rest-frame) luminosity of the host as derived
from the measured colours would also contribute to the total $K$-band
QSO luminosity about 5\%.  Our measurements indicate large and
luminous extended systems ($D_{eff}\approx 4$~Kpc, $R\approx21$ to 22~mag).  
We will consider three alternative
explanations for their nature: scattered light from the active
nucleus, nebular continuum produced by the gas ionized by the active
nucleus, and stellar light.

\subsection{Scattering}
The large structures detected in high-redshift radio-galaxies
(e.g. Chambers et al. 1987) have been attributed to light scattered 
from the active nucleus by a powerful transverse radio-jet
(e.g. Fabian 1989).  Indeed, recent spectro-polarimetry of two
$z\approx1$ radio-galaxies that exhibit 'alignment effects' (optical
structures oriented in the direction of the radio-jet) indicates that
around 80\%\ of the total UV continuum emission at rest-frame
\ldo{2800} is non-stellar scattered light (Cimatti et al. 1997).
Since radio-galaxies and radio-loud QSOs could be identical objects
viewed from different angles (Barthel 1989), an important amount of
scattered light may be present around radio-loud nuclei (see Fosbury
1997 for a detailed discussion).

The host of the core-dominated radio-loud QSO studied in this paper,
PKS 2134$+$008, has an $R-I$ colour that is 0.52~mag redder than the
nucleus. This argues against the simple optically-thin scattering
case, which would yield colours as blue as or bluer than the nucleus
itself. 
 The $R-K$ limits for this host ($\lsim 3.3$~mag) are not deep
enough to probe the optically-thin scattering case value.  Note that
four of the six lobe-dominated radio-loud QSOs at $z\approx 2-2.5$
studied by Lehnert et al.\  (1992) have hosts with $B-K$ colours redder
than the nuclei.  However the $B-K$ (or $R-K$)
analysis discriminates poorly the
origin of the light below the \ldo{4000} break, since scattered 
light would contribute predominantly in $B$-band (rest frame
\ldo{1400}) 
and stellar light  would contribute predominantly in  $K$-band
(rest-frame \ldo{7300}) 
(e.g. Fosbury 1997).

The host of our radio-quiet QSO with calibrated photometry,
1630.5$+$3749, exhibits the same general properties, with colours 0.25
and 0.8~mag redder than the nucleus in $R-I$ and $R-K$, respectively.
These colours
are inconsistent with the optically-thin scattering case as an
explanation for the UV hosts of our QSOs. An optically-thick medium
should be invoked in order to produce colours redder than those of the
scattered source, but then the geometry of the scatterers should not
be symmetrical since we still see the blue colours of the nucleus
itself. This could be an alternative mechanism for the origin of the
extended light in our sample, although there is no evidence in general
for asymmetric scattering in radio-quiet QSOs.

\subsection{Nebular light}

  In ABT95 we argued that nebular continuum produced by an extended
narrow line region around the active nucleus was unlikely to be the
origin of the hosts we have detected, the reason being that the
expected narrow emission lines should be very prominent in that case,
with peak intensities of the Balmer lines more than 3 times larger
than those of the broad lines. We have since then acquired near-IR
spectra of two of the QSOs studied here (see Fig.6) which show
prominent broad \Ha\ lines, but no prominent narrow components. Note
that even \hbox{[O III]\ldo{5007}}, marked in the figure, is
marginally detected at best 

\begin{figure*}
    \cidfig{5in}{31}{77}{584}{725}{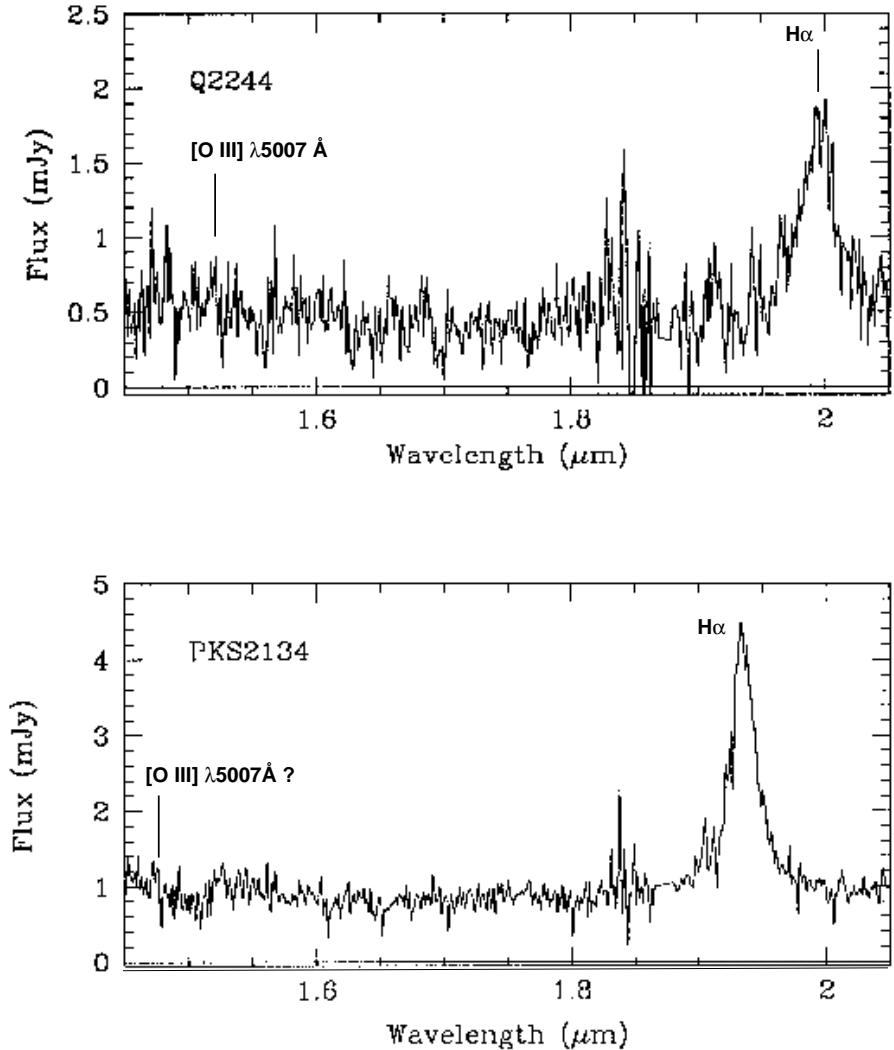}
\caption{ Near-IR spectrum of the QSOs PKS 2134$+$008 and Q 2244$-$0105
}
\end{figure*}

\subsection{Stellar light}

Stellar light remains still the most plausible interpretation for the
extended light we have detected. The colours, sizes, luminosities and
radial profiles are indeed in agreement with those expected from young
stellar populations:

a) The {colours} of the hosts, $R-K \approx 3.3$~mag for 1630.5$+$3749
and $R-K \lsim 3.3$ for PKS 2134$+$008 and Q 2244$-$0105, are not
consistent
with the colours predicted from a simple passively evolved stellar population,
usually assumed to be characteristic of
elliptical galaxies. Bressan, Chiosi \& Fagotto (1994), for instance, calculate
that a passively-evolved stellar population for a $z\sim2$ 
elliptical galaxy be $R-K \approx5.5$~mag and slightly bluer for a mild 
amount of activity. Younger populations are necessary in order to 
account for the blue colours observed in our hosts. As an example,
typical nearby H~II galaxies set at $z=2$
would have colours as blue as $R-K \sim 1-1.7$,
as derived from their characteristic 
flat spectral energy distributions (SED): $f_\nu \propto \nu^\alpha$, 
with $0 \lsim \alpha \lsim 0.5$.

b) and c) The {luminosities and radii} of our hosts lie along the 
luminosity--radius relationship of the local young H~II galaxies 
(Telles, Melnick \& Terlevich 1997).
In Fig.~7  we converted the UV luminosities
of the hosts (observed $R$-band) to rest-frame $B$-band
using the SED of local H~II galaxies.
This is equivalent to converting the $B$-band luminosities of 
the local sample of H~II galaxies to rest-frame \ldo{2300}, 
and then comparing them with those of the hosts. 
Note that there is
at least one local H~II galaxy that is as big and luminous as our hosts.

\begin{figure*}
    \cidfig{2.8in}{24}{154}{293}{426}{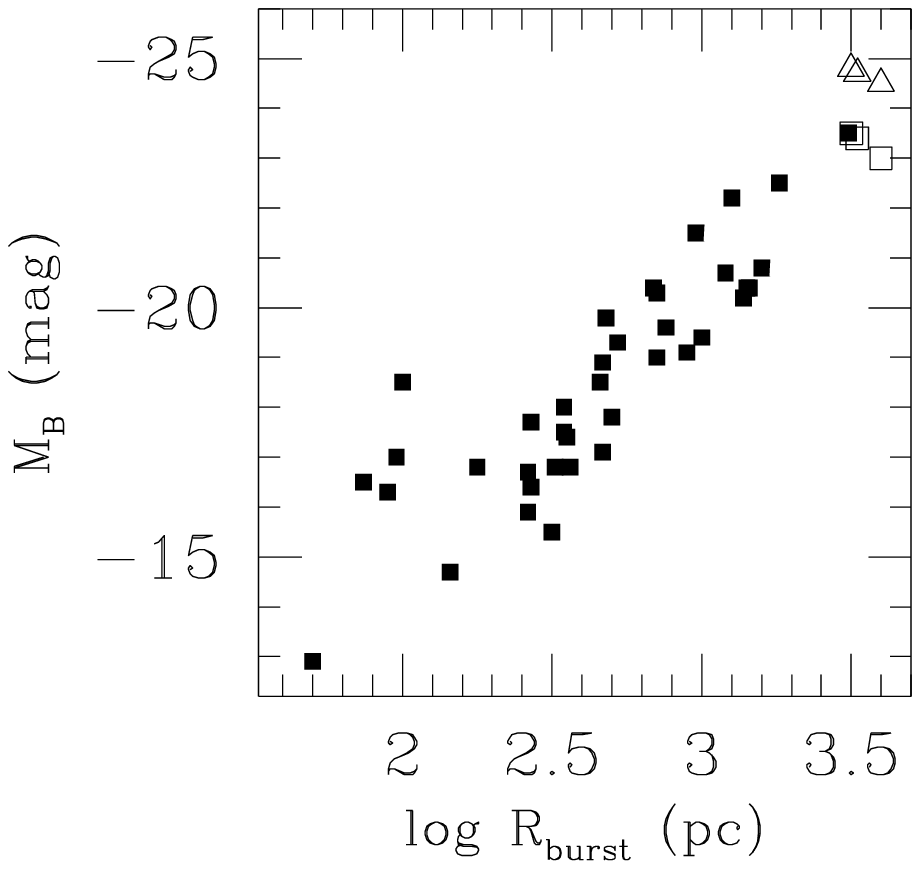}
\caption{Luminosity--size relationship for nearby H~II galaxies
and QSOs hosts. H~II galaxies are marked with filled squares.
The three QSOs studied here lie in this relation 
if their SEDs are $f_\nu \propto \nu^0$ (open triangles) to 
$f_\nu \propto \nu^{0.5}$ (open squares). These SEDs are 
typical of young H~II galaxies}
\end{figure*}

d) The radial profiles of 
the $R$-band hosts, derived from the flat-top solutions, 
fall approximately as $r^{1/4}$-laws or exponential profiles for radii 
$r \gsim 0.6$~arcsec (Fig. 8).
Profiles derived for radii smaller than the FWHM of the observations 
are usually unreliably 
recovered by flat-top subtractions, as shown by our numerical
simulations of galaxy$+$PSF (see section 3.1.2). Distinguishing between 
different morphologies is a hopeless task at these redshifts
since the discrimination between disks and bulges is most efficiently
carried out in the the inner part of the profiles, where our data
are unreliable. The point we want to illustrate with Fig.~8 is that
the morphology of the hosts is not definitely different from that of 
smooth local galaxies, at least in the outer regions.

\begin{figure*}
    \cidfig{5in}{36}{394}{577}{694}{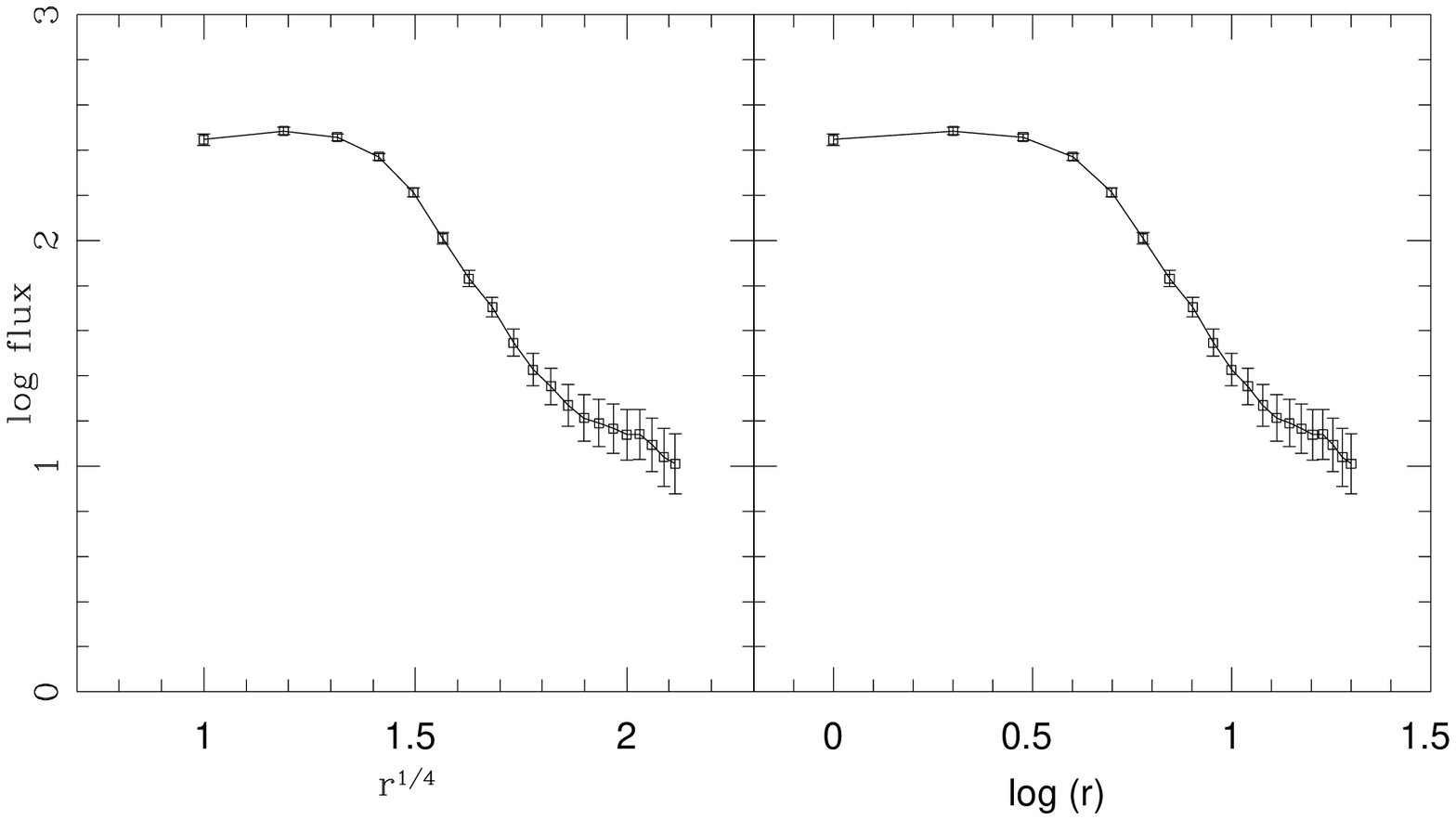}
\caption{ Radial profile of the $R$-band host of the QSO 1630.5$+$3749,
in a log counts vs. $r^{1/4}$ and in a log$(r)$ diagram. Elliptical
profiles and disk-like profiles 
give linear relationships in 
the right-hand and left-hand panel, respectively.}
\end{figure*}

\ifoldfss
  \section{Conclusions}
\else
  \section[]{Conclusions}
\fi

We have shown that the hosts of $z\approx2$ high-luminosity
radio-quiet and radio-loud QSOs are large and luminous, rivaling the
most luminous nearby star forming galaxies. Colours, sizes,
luminosities and radial profiles are compatible with the
interpretation of the extended light as a young galaxy.  If all the UV
luminosity is coming from a stellar population, the implied star
formation rates (in the continuous star formation case) are SFR$\gsim
100-200$~\Msun/yr on scales of $D_{eff}\approx 4$~Kpc. These 
values are about an order of magnitude higher than those derived for field
galaxies at similar redshifts found by Lyman Break searches in the
Hubble Deep Field (Steidel et al. 1996b, Lowenthal et al. 1997).  At
$z\approx 2$, an unevolved $L_\star$ galaxy with SED typical of a
star-forming galaxy would appear to be about 3 mag fainter than the
hosts we have detected.  The density of high-luminosity QSOs ($M_B
\lsim -28$~mag) at redshifts between $1\lsim z \lsim 3$ (Boyle et
al. 1991)
is about 10 Gpc$^{-3}$.  This means that the density of large luminous
galaxies like the ones considered here should be about 8 Gpc$^{-3}$ at
$1<z<3$, taking into account the non-detection case of ABT95 and the
relative under-abundance of radio-loud QSOs, which form less than 10\%
of the QSO population.

While these results are intriguing, it is still possible that other
sources such as nebular light and scattering may contribute to the
observed extensions.  However, the absence of prominent narrow
emission lines in the spectra of these QSOs argues against a major
contribution by the nebular continuum emitted by an extended narrow
line region. The optically-thin scattering hypothesis is furthermore
ruled out as a major mechanism by redder UV and UV-optical 
colours than those of the
active nuclei. The optically-thick scattering case is still an open
possibility, especially in radio-loud hosts, where a role for
scattering has already been invoked.  High spatial resolution
spectroscopic observations of the extended light is probably the
simplest and most direct way to assess the contribution of all those
different mechanisms into the UV luminosities we have measured. This
is an important project in order to reliably prove and measure the
contribution of massive star-formation in QSO hosts.

The study of the hosts of less luminous QSOs, especially those
which populate the break of the QSO luminosity function, is also of
major importance, in order to assess the association of large luminous
galaxies and the majority of QSOs at the QSO epoch.

\section*{Acknowledgments}
We thank Eugene Churazov, Robert Fosbury and Simon White for useful 
discussions, and James Lowenthal for also providing comments on
an early draft of this paper. We thank Tom Geballe and Sandy 
Leggett of the UKIRT service-time team for tenaciously tracing down 
the IR spectra in the archives after we lost them in a disk crash. 
This work was supported in part by the `Formation and Evolution of
Galaxies' network set up by the European Commission under contract 
ERB FMRX-CT96-086 of its TMR programme.

\end{document}